\documentclass[prd,aps,twocolumn]{revtex4}
\usepackage{bm}
\usepackage{graphicx} 
\usepackage{amsmath,amssymb}
\usepackage{aas_macros}
\usepackage{color}
\usepackage{ulem}
\usepackage{here}
\input{colordvi.tex}

\begin{document}
\date{\today} 
\title{Effects of axions on Nucleosynthesis in massive stars} 

\author{Shohei Aoyama}
\affiliation{Department of Physics and Astrophysics, Nagoya University,
Nagoya 464-8602, Japan} 

\author{Takeru K. Suzuki} 
\affiliation{Department of
Physics and Astrophysics, Nagoya University, Nagoya 464-8602, Japan}

\begin{abstract}
 We investigate the effect of the axion cooling on the nucleosynthesis
 in a massive star with $16M_{\odot}$ by standard stellar evolution
 calculation. We find that the axion cooling suppresses the nuclear
 reactions in carbon, oxygen and silicon burning phases because of the
 extraction of the energy. 
 As a result, larger amounts of the already synthesized neon
 and magnesium remain without being consumed to produce further heavier elements. 
 Even in the case with the axion-photon coupling
 constant $g_{a\gamma }= 10^{-11}$ GeV$^{-1}$, which is six times smaller than the current upper
 limit, the amount of neon and magnesium that remain just before 
the core-collapse supernova explosion is considerably larger than the standard value.
This implies that we could give a more stringent constraint on $g_{a\gamma }$ from the 
nucleosynthesis of heavy elements in massive stars. 
\end{abstract}

\maketitle

\section{Introduction}
The standard model of particle physics (SM) 
well explains general properties of the results of collider experiments.
For example, the lattice QCD calculation,
which is the simulation based on the first principle of the quantum chromodynamics (QCD), 
can predict the mass of baryons and mesons only from a few input parameters. 
{On the other hand,} the Lagrangian density of QCD has a term which violates the CP symmetry and 
involves the finite electric dipole moment (EDM) of neutron. 
Because this EDM has never {been} detected, 
{it is generally considered}
that QCD has a fine tuning problem named the strong CP problem. 
Peccei and Quinn suggested that the existence of an undiscovered
pseudo-scalar particle which is associated with another $U_{\rm A}(1)$ symmetry for SM 
can solve this problem \cite{PhysRevLett.38.1440}. 
This pseudo-scalar particle is named axion and have interactions 
with baryons, leptons and photons 
({\it see. e.g.} \cite{2008LNP...741.....K,1990eaun.book.....K,1996slfp.book.....R}).
The coupling constant, $g_{a\gamma }$, of 
axions to photons is
related to the energy scale of the symmetry breaking $f_{a}$ as
\begin{equation} 
g_{a\gamma }=\dfrac{\alpha C_{\gamma }}{2\pi f_{a}}~,
\end{equation} 
where $C_{\gamma }$ is a model-dependent constant. 
For KSVZ \cite{1979PhRvL..43..103K,1980NuPhB.166..493S} 
and DFSZ \cite{1981PhLB..104..199D} scenarios, 
$|C_{\gamma }|=1.9$, and $0.7$, are adopted, 
respectively, and several constraints on them 
have been set ({\it e.g.} \cite{2008LNP...741.....K,1996slfp.book.....R}).
In addition, axions, which have a finite mass 
as a results of the symmetry breaking, 
are a candidate of {the} cold dark matter
({\it e.g.} \cite{1990eaun.book.....K}).

The interaction of axions with photons is supposed 
to affect the structure and the evolution of a star.
Because the predicted mass of {an axion} is smaller 
than the typical temperature of the stellar interior,
axions are expected to be easily produced in 
stars through the interaction with photons
\footnote{Electrons, heavy ions and nucleons 
in stars are also expected to play an 
important role in the axion production process.
{However we focus only on axions generated by photons.}}. 
The conversion from photons to 
axions removes the heat in the stellar interior, which possibly gives an impact on the 
stellar structure, whereas its reaction rate strongly 
depends on temperature. 

Various possibilities concerning axions in the stellar interior
have been explored in a wide range of stellar mass.
By comparing the photon luminosity of the sun with 
the nuclear reaction rate that is calibrated from 
the neutrino luminosity, Gondolo and Raffelt set a constraint 
$g_{a\gamma }<7\times 10^{-10}$~GeV$^{-1}$ \cite{2009PhRvD..79j7301G}. 
Tighter constraints can be obtained for stars in later 
evolutionary stages because the temperature in the interior is higher than 
the temperature in main sequence stars, {\it e.g.} the sun, and the production 
rate of axions is larger as well. 
For example, a constraint is derived from number counts of horizontal branch 
stars; the generation of axions tends to shorten the duration of the horizontal 
branch phase,  {which contradicts to} the standard stellar 
model {without the effect of} axions 
{that} reproduces the observed distribution of horizontal branch stars within 
10 \% accuracy. 
From this observational requirement, 
Ayala gives $g_{a\gamma }<0.66\times 10^{-10}$~GeV$^{-1}$\cite{2014arXiv1406.6053A}. 
Massive stars also give tight constraints on $g_{a\gamma }$, since 
the interior temperature is suitable for the generation of axions. 
If one takes into account axions, the duration of the Helium 
burning is shorten, which would erase the blue loop stage in the 
Hertzsprung-Russell (HR) diagram required for observed Cepheid variable stars. 
By considering this effect for stars with 8 - 12 $M_{\odot}$, 
where $M_{\odot}$ is the solar mass, Friedland {\it et al.} found 
$g_{a\gamma }<0.8\times 10^{-10}$~GeV$^{-1}$ \cite{2013PhRvL.110f1101F} 
with {\tt MESA} \cite{2011ApJS..192....3P,2013ApJS..208....4P},  
a public code for one-dimensional calculations of stellar evolution.

In this paper we consider the effect of axions in more 
massive stars. Pantziris \& Kang estimated a constraint on the axion cooling 
rate {for such massive stars by using a simple one-zone model instead of 
realistic stellar structure} \cite{1986PhRvD..33.3509P}. 
In contrast, we focus on the effect of axions on the nucleosynthesis of 
heavy nuclei at the very late phase of the stellar 
evolution just before the core-collapse supernovae. 
Heavy elements such as silicon, 
sulfur and iron are synthesized during the last few months 
({\it see.} \cite{2010sen..book.....R}).
Because of the considerably short duration of the nucleosynthesis 
of these elements, the photons generated through the nuclear reactions, which 
do not have enough time to travel to the stellar surface, can be 
hardly observed.
A part of the synthesized heavy nuclei are finally ejected and affects 
elemental abundances of next generation stars and the chemical evolution of 
galaxies. The produced amounts of these 
heavy metals have the information 
of the high-temperature and high-density environment
of the stellar interior.

\begin{figure}
 \begin{center}
  \includegraphics[width=60mm]{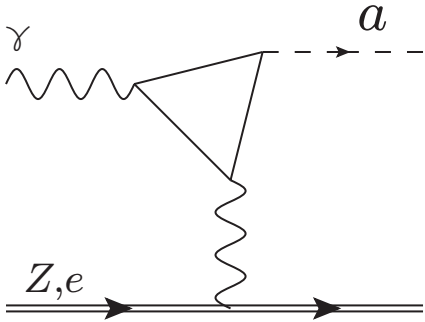}
 \end{center}
  \caption{The Feynman diagram of the Primakoff process. 
$Z,~e$ represent ions and the electron which provide 
a photon via the magnetic field.
}  \label{fig.primakoff}
\end{figure}

In this study, we focus on the Primakoff process \cite{PhysRev.81.899}, 
namely axions produced by the conversion from two photons (figure \ref{fig.primakoff}),  and 
estimate the effects of axion cooling 
on the nucleosynthesis at the 
late phase of massive stars with the {\tt MESA} code.  
This paper is organized as follows. 
Our treatment of the axion cooling in the stellar evolution 
code is shown in \S 2. 
In \S 3, we compare the structure of 
a massive star with axions to 
a standard case that does not take into account axions.
In \S 4, we discuss the effect of the axion 
cooling on the explosive nucleosynthesis.
\S 5 concludes this paper. 

\section{Setup}
\subsection{Axion Cooling}
The axion cooling rate per unit mass $\varepsilon_{a}$ in hot plasma 
has been derived by several authors 
({\it e.g.} \cite{1982PhRvD..26.1840F,1986PhRvD..33..897R,1986PhRvD..33.3509P,1987PhRvD..36.2211R,1990PhR...198....1R,1990ZPhyC..47..559A,1994APh.....2..175A}).
They showed that the Primakoff process plays a primary role in the axion cooling 
when electrons in stars are 
non-relativistic. 
In particular, if the plasma frequency $\omega_{0}$ is small enough to satisfy 
the condition, $h \omega_0\ll k_{\rm B}T$, besides the non-relativistic and 
non-degenerate condition fulfilled, 
the emission rate can be obtained analytically as \cite{1990PhR...198....1R,1996slfp.book.....R}
\begin{equation} 
\varepsilon_{a({\rm NR\& ND})}=\dfrac{g_{a\gamma }^{2}T^{7}}{4\pi \rho}F(k_{\rm S},T)
\end{equation}
where $k_{\rm S}$ is the Debye-H\"{u}ckel wave number which {is} defined 
in the Raffelt \cite{1988PhRvD..37.1356R,1990PhR...198....1R,1996slfp.book.....R}.
The function $F(k_{\rm S},T)$ is defined as \cite{1986PhRvD..33..897R,1990PhR...198....1R,1996slfp.book.....R}
\begin{equation} 
F(k_{\rm S},T)=\dfrac{\kappa^{2}}{2\pi^{2}}
\displaystyle\int_{0}^{+\infty}\!\!\!\!\!\!\!
\left(x^{2}+\kappa^{2}\right)\ln \left(1+\dfrac{x^{2}}{\kappa^{2}} \right)
\dfrac{x}{e^{x}-1}dx~,
\end{equation}
where $x \equiv \hbar \omega \slash k_{\rm B}T$ and $\kappa =2c k_{\rm S}\slash \hbar k_{\rm B}T$ \footnote{$\hbar$, $k_{\rm B}$ and $c$ are the reduced Planck constant, the Boltzmann constant and the speed of light, respectively.}. 
In horizontal branch stars and Oxygen burning stars, $\kappa^{2} =2.5,~$ 
In the region where the radiation pressure {dominates}, since 
the radiation pressure is proportional to $T^{4}$, one can find \cite{2013sse..book.....K}
\begin{equation} 
\dfrac{T}{\rho^{1\slash 3}}=\left(\dfrac{3c\mathcal{R}}{4\sigma \mu}\right)^{1\slash 3}~,
\end{equation}
where $\mathcal{R}$ and $\sigma $ are the gas constant and the Stefan-Boltzmann constant, respectively. 
Because the value of {the} right hand side is almost stationary in the stellar evolution,
one can regard ${T^{3}}\slash {\rho}$ as a constant and find that 
$\varepsilon_{a({\rm NR\& ND})} \propto T^{4}$. 
This dependence {is} mentioned by several authors ({\it e.g.} \cite{1982PhRvD..26.1840F,1986PhRvD..33..897R,1986PhRvD..33.3509P}).

On the other hand, when one considers the nucleosynthesis of heavy elements, 
the temperature is so high that the relativistic effect needs to 
be taken into account, although the electrons are still non-degenerate 
\cite{2013sse..book.....K,1994APh.....2..175A}. 
Altherr {\it et al.} reported that
the formula which is valid in the limit of the non-relativistic and
non-degenerate plasma 
can be used to the relativistic plasma, whose temperature exceeds the rest mass of an
electron $m_{\rm e}$, {\it i.e.} $k_{\rm B}T \gg m_{\rm e}c^{2}$ 
\cite{1990ZPhyC..47..559A}
In addition, as the plasma frequency $\omega_{0}$ increases, the emission rate 
suffers an exponential cut-off, $\propto \exp (-\hbar \omega_{0}\slash k_{\rm B} T)$ 
({\it see.}\cite{1982PhRvD..26.1840F}). 
Hence we adopt the following formula for the axion cooling rate 
\begin{eqnarray} 
\label{eq:axioncooling}
\varepsilon_{a} &=&\varepsilon_{a({\rm NR\& ND})}
\exp \left(-\frac{\hbar \omega_{0}}{ k_{\rm B} T}\right)\\
&=&27.2 g_{a\gamma 10}^{2}T_{8}^{7}\rho_{3}^{-1}F(k_{\rm S},T)\label{primakoff}
\exp\left(-\dfrac{\hbar \omega_{0}}{k_{\rm B}T}\right)~[{\rm erg/g/sec}]~,\notag
\end{eqnarray}
where $g_{a\gamma 10}\equiv g_{a\gamma }/10^{-10}$ GeV$^{-1}$. 
$T_{8}$ and $\rho_{3}$ are temperature normalized by $10^{8}$ K and 
density normalized by $10^{3}$ g/cm$^{3}$, respectively. 

\subsection{Stellar Evolution}
We include the cooling by axions, eq.(\ref{eq:axioncooling}), 
in the stellar evolution code {\tt MESA}. We add the extra term for the axion 
cooling in the energy transfer equation that is one of the basic equations 
governing the evolution of the stellar structure. 
The axion cooling term simply removes the luminosity carried by the photons 
emitted as a result of the nuclear reactions in the stellar interior. 
Therefore, it reduces the radiation pressure by these photons to modify 
the momentum balance, and accordingly, changes the stellar structure if 
the effect is not negligible. 

We calculate the evolution of a star with 
$M=16 M_{\odot}$, with the solar elemental abundances 
from the zero-age main sequence phase when the hydrogen burning reaction 
is ignited at the center of the star. 
We take into account the mass loss by radiation pressure-driven stellar wind 
with an empirical mass loss rate \cite{1975MSRSL...8..369R}. 
We follow the time evolution until the gravitational 
core-collapse sets in just before the supernova explosion.
In addition to the cases with the axion cooling, we also calculate the 
standard case that does not include the effect of axions for comparison.

\section{Results}
\subsection{Overview of Stellar Evolution}
A star changes its luminosity and surface temperature with time. After the 
exhaustion of the hydrogen in the central core, a star evolves to a red giant 
and eventually the helium burning sets in to synthesize heavier elements. 
Massive stars with $M\gtrsim 10M_{\odot}$ continues through the oxygen 
burning to the silicon burning phase with their core being non-degenerated, 
iron group elements dominate the core finally before the core-collapse 
supernova ({\it see.} \cite{2010sen..book.....R}). 
With the stellar evolution, the star with the initial mass of 
16 $M_{\odot}$ lost $\approx 2M_{\odot}$ during its lifetime
by radiation 
pressure-driven stellar wind ({\it see.} \cite{2013sse..book.....K}).

\begin{figure}
 \begin{center}
  \includegraphics[width=80mm]{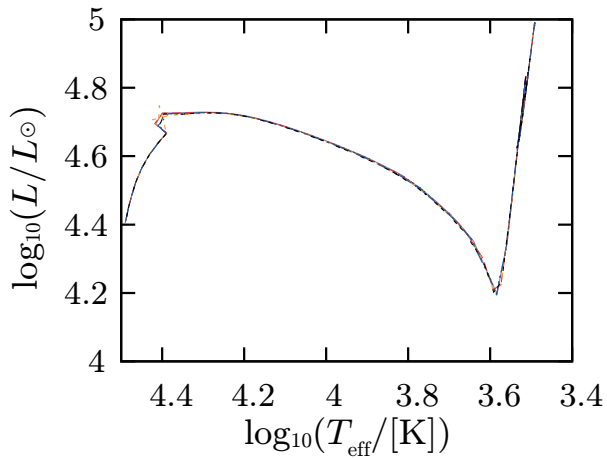}
 \end{center}
  \caption{Evolutionary tracks of stars with $16 M_{\odot} $ that take into 
    account axion coolings with different coupling constants, $g_{a\gamma }$, 
    in an HR diagram. 
 The long dashed-dashed (brown), long-dashed (red) and dashed
 (orange) represent the evolutionary track with
 $g_{a\gamma }=10^{-10},~10^{-11},$ and $10^{-12}~\mathrm{GeV}^{-1}$,
 respectively. The solid (purple) line indicates the case without axion
 cooling.
 }
 \label{fig.hr}
\end{figure}

We study the effect of the axion cooling in the Hertzsprung-Russell (HR) diagram. In figure \ref{fig.hr}, we plot evolutionary
tracks in cases with $g_{a\gamma }=10^{-10}$ and $10^{-11}~\mathrm{GeV}^{-1}$ 
in comparison with the case without axion cooling
\footnote{Although $g_{a\gamma}=10^{-10}~\mathrm{GeV}^{-1}$ has already been excluded in a
number of observations ({\it e.g.}
\cite{2011ApJS..192....3P,2013ApJS..208....4P}),
we plot this case for an illustrative purpose, 
in order to show the effect of the axion cooling.
}.
One can find that 
these three cases show almost no difference. 
The main reason is that the duration of the carbon, oxygen and 
silicon burning phases, 
which are largely affected by the axion cooling, {is $\sim$ several thousands years. This} 
is much shorter than that of the thermal time scale of the star
($\sim$ a hundred thousand years.) \cite{2013sse..book.....K}.
Namely, the effect of the nuclear reaction deeply in the stellar interior 
is not still observable because the travel time of the photons to the stellar 
surface is much longer. Therefore, the effect of the axion cannot be 
observed in the HR diagram.

\subsection{Nucleosynthesis }\label{axion.structure}
We investigate how the axion cooling affects the nucleosynthesis 
through the evolution of the massive star. 

\begin{figure}
 \begin{center}
  \includegraphics[width=60mm]{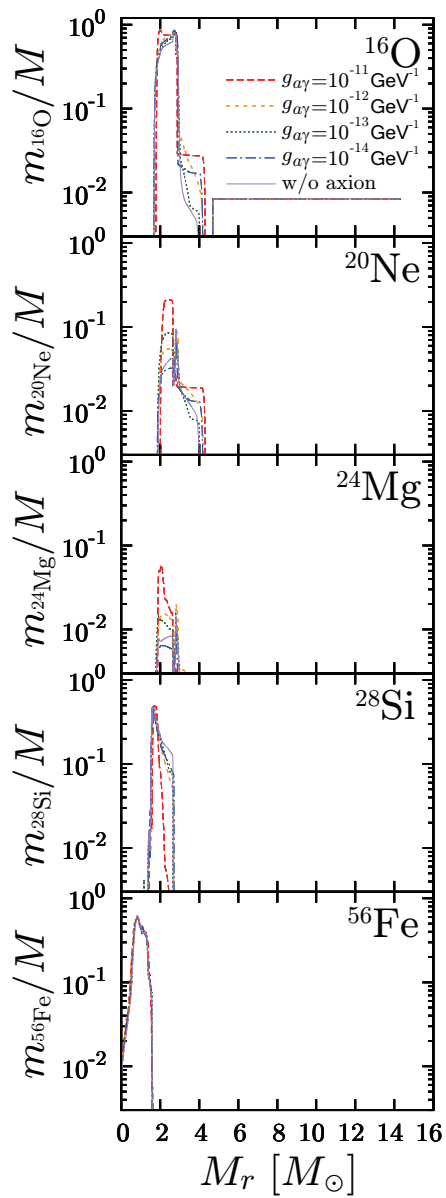}
 \end{center}
  \caption{The abundance of alpha elements as a function of mass radius in 
    units of the solar mass, 
    $M_r/M_{\odot}$, at the end of the silicon burning 
    phase. 
Long-dashed (red), dashed (orange), dotted (dark-green), 
dot-dashed (blue) lines 
correspond to the cases with
 $g_{a\gamma }=10^{-11},~10^{-12},~10^{-13}$, and $10^{-14}$, respectively.
Solid line represent the case without axion cooling. 
}  \label{fig.16M}
\end{figure}

We compare abundances of alpha elements, oxygen 
($^{16}$O), neon ($^{20}$Ne), magnesium ($^{24}$Mg), and silicon ($^{28}$Si), 
in addition to iron ($^{56}$Fe) of the star with different $g_{a\gamma }$ 
and initial mass of $M=16~M_{\odot}$ in figure \ref{fig.16M}.
The abundance of $^{20}$Ne and $^{24}$Mg are enhanced for large $g_{a\gamma }$ 
near $M_{r}\simeq 2M_{\odot}$, whereas the behaviors are complicated. 
On the other hand, the abundance of $^{28}$Si shows the opposite trend.  
The produced amount of $^{56}$Fe 
is not influenced significantly by the axion cooling even with 
large $g_{a\gamma }\geq 10^{-12}~{\rm GeV}^{-1}$.

\begin{figure}[H]
 \begin{center}\label{figure5}
  \includegraphics[width=80mm]{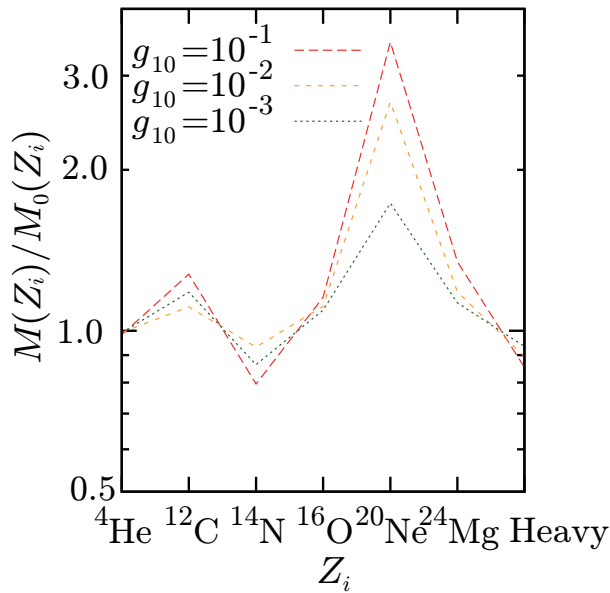}
 \end{center}
  \caption{The abundance of each element $i$ ($M(Z_{i})$) normalized by that 
    from the standard case without axion cooling ($M_{0}(Z_{i})$)
just before the core collapse. 
Long-dashed (red), dashed (orange), and dotted (dark-green) lines
correspond to the cases with
 $g_{a\gamma }=10^{-11},~10^{-12}$, and $10^{-13}$  GeV$^{-1}$, respectively, 
where $g_{10}\equiv g_{a\gamma }\slash (10^{-10} $GeV$^{-1})$.
"Heavy" means the total amount from $^{28}$Si to
 $^{56}$Ni. } \label{fig.temp}
\end{figure}

In figure \ref{fig.temp}, 
we present the total mass of each element left just before the core collapse 
{(type II) supernova, 
where each value is normalized by that of the standard case without axion cooling.}
Metals heavier than $^{24}$Mg are converted to iron-group elements during 
the explosive nucleosynthesis just after the explosion 
\cite{1995ApJS..101..181W}. 
Hence, as for the heavy elements from $^{28}$Si to $^{56}$Ni, 
we consider the total amount of them.

The amount of $^{20}$Ne is considerably increased for large $g_{a\gamma }$; 
in the case with $g_{a\gamma }=10^{-11}$ the amount is enhanced more than 
three times. This trend is also weakly seen in the amount of $^{24}$Mg. 
On the other hand, the total amount of the heavier elements is smaller 
for larger $g_{a\gamma }$.

\begin{figure}
 \begin{center}
  \includegraphics[width=80mm]{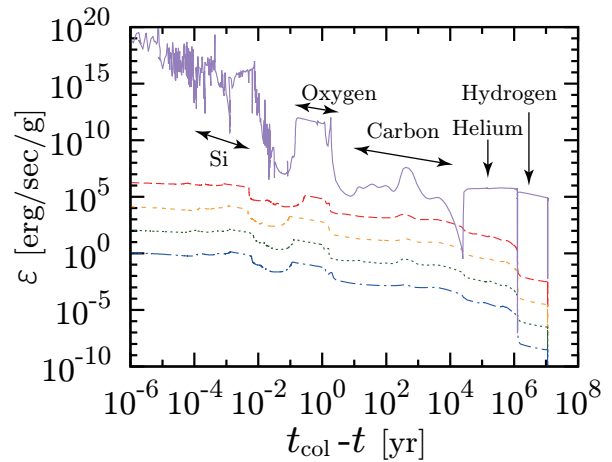}
 \end{center}
  \caption{The time evolution of axion emission rate per unit mass from
 ZAMS (right side at $t_{\rm col} - t =10^7$ yr) to core-collapse (left side to 
 $t_{\rm col} - t \Rightarrow 0$) with 16 $M_{\odot}$, where $t_{\rm col}$ 
 is the time of the core collapse.. Long-dashed (red), 
 dashed (orange), dotted (dark-green), 
 dot-dashed (blue) lines represent the cases
 with $g_{a\gamma }=10^{-11},~10^{-12},~10^{-13}$, and $10^{-14}$,
 respectively. 
Solid (purple) line represents the energy generation rate due to the nuclear
 reactions {from the standard case} without axion cooling. 
}  \label{fig.time-eps}
\end{figure}

To understand these results, we plot in figure \ref{fig.time-eps} the 
time evolution of both the total energy generation rate and 
the axion emission rate at the center
from the ZAMS to the core collapse.
In the weak coupling cases
{with} $g_{a\gamma }\lesssim 10^{-10}$ GeV$^{-1}$ 
{we are considering}, the axion cooling does not 
change the lifetime of the star. 
{In such circumstances,} the effect of axion cooling 
is {simply the} extraction of {the} thermal energy 
of the star, {which slows down the} nuclear 
reactions\footnote{
{In general cooling increases the temperature of a star because the 
gravo-thermal specific heat is negative. }
However, in this case, 
{the duration of the nuclear burning}, $\mathcal{O}(10^{3})$ years, 
 is much {shorter} than 
Kelvin-Helmholtz time scale of the star, $\mathcal{O}(10^{5})$ years.
{Therefore}, the axion cooling of the star is too 
rapid to change the inner structure of the star, and invokes
the decrease of {the} temperature of the star.}.
{Our calculation shows} that the temperature {at the central region} with 
$g_{a\gamma }=10^{-11}$ GeV$^{-1}$
is 1 \% lower than the standard case without axion cooling  
at the time of the termination of the Silicon burning at the center.
In general, nuclear reaction rates strongly depend on  temperature
({\it e.g.} \cite{2010sen..book.....R,2013sse..book.....K}) and
{the small decrease of} the temperature {considerably slows down the} nuclear reactions.
In fact, {the peak value of the nuclear reaction rate is suppressed} by half due to the axion
cooling at that time. 

One can find in figure \ref{fig.time-eps} that the axion cooling rate 
increases monotonically over time at the {central} region.
Therefore, the nuclear reactions in later phases such as carbon, oxygen and silicon burning phases
are suppressed by the axion cooling and 
the larger amount of $^{20}$Ne and $^{24}$Mg is left, avoiding nuclear burning to 
synthesize heavier elements.
Hence, the abundance of these elements, which are synthesized during these 
phases, are enhanced by the axion cooling as a result of {the} inactivation
of the nucleosynthesis. Since these elements are not so affected 
by the later explosive nucleosynthesis \cite{1995ApJS..101..181W}, 
the ejected mass would be also larger for larger $g_{a\gamma }$ to possibly 
affect the chemical evolution of galaxies.

\section{conclusion}
In this paper, we have studied the effect of the axion cooling
on the nucleosynthesis in the massive star with $16 M_{\odot}$
by using the stellar evolution code, {\tt MESA}.
We have found that the axion cooling suppresses the nucleosynthesis 
in the carbon, oxygen and silicon burning phases
even in the weak coupling, $g_{a\gamma }<10^{-11}$ GeV$^{-1}$. 
As a result, the abundances of oxygen, neon and magnesium
 increase as the coupling constant
$g_{a\gamma }$ increases.
Even in the case $g_{a\gamma}=10^{-11}$ GeV$^{-1}$, which is six times
smaller than the current upper limit of $g_{a\gamma}=10^{-10}$,
the final amount of neon is enhanced more than three times larger than the
standard value that does not take into account the effect of axion cooling.

\section{Acknowledgment}
This work is supported in part by scientific research grant
for Research Fellow of the Japan Society
for the Promotion of Science from JSPS Nos. 24009838
(SA). 

\bibliography{reference2013f_suzukitakeru}

\end{document}